\pdfoutput=1 
\documentclass[fleqn,usenatbib,useAMS]{mnras}
\usepackage{amssymb}
\usepackage{amstext}
\usepackage{amsmath}
\usepackage{epsf}
\usepackage{graphicx}
\usepackage{graphics}
\usepackage{epsfig}
\usepackage{wrapfig}
\usepackage{listings}
\usepackage{float}
\usepackage{esint}
\usepackage{natbib}
\usepackage{ifthen}
\usepackage{hyperref}
\usepackage{verbatim}
\usepackage{xcolor}
\usepackage[normalem]{ulem} 
\newcommand\hl{\bgroup\markoverwith
  {\textcolor{yellow}{\rule[-.5ex]{2pt}{2.5ex}}}\ULon}

\newcommand\rhl{\bgroup\markoverwith
  {\textcolor{orange}{\rule[-.5ex]{2pt}{2.5ex}}}\ULon}
  
 \newcommand\ghl{\bgroup\markoverwith
  {\textcolor{green}{\rule[-.5ex]{2pt}{2.5ex}}}\ULon} 
\begin{document}

\title[Caustics in the Dark Matter Sheet]{The Caustic Design of the Dark Matter Web}

\author[Shandarin \& Ramachandra]
	{Sergei F. Shandarin$^1$\thanks{E-mail: sergei@ku.edu}, 
	Nesar S. Ramachandra$^2$\thanks{E-mail: nramachandra@anl.gov} \\
	$^1$Department of Physics and Astronomy, University of Kansas, Lawrence, KS 66045 \\
	$^2$High Energy Physics Division, Argonne National Laboratory, Lemont, IL 60439
	}

\maketitle

\begin{abstract}

Matter density is formally infinite at the location of caustic surfaces, where dark matter sheet folds in phase-space. 
The caustics separate regions with different number  of streams and the volume elements change the parity  by turning inside out
when passing through the caustic stage. 
Being measure-zero structures, identification of caustics via matter density fields is usually restricted to fine-grained simulations.
Here we employ a generic algorithm to identify caustics directly using the triangulation of the Lagrangian sub-manifold ${\bf x}({\bf q},t)$
obtained in N-body simulations.
In our approach the caustic surfaces are approximated by a set of triangles whose vertices are particles  of the simulation.  
The major obstacle we encountered was insufficient sampling of small scale perturbations.
We overcame it by a brute force approach. We continued to raise the scale of the cutoff in the initial power spectrum until obtained
the reliable resolution of the caustics shells up to seven layers. Although quite modest, our result is the first reliable direct   
construction of caustic surfaces in N-body simulation. It reveals a number of unexpected geometrical features. 
In particular shapes of some of them are contrastingly different from the known shapes of the caustics formed in the Zeldovich approximation. 

\end{abstract}



\section{Introduction}
Caustics along with the multi-stream and flip-flop fields are  inherent features of the cold collision-less 
dark matter (DM) web.
Known in geometric optics as an  envelope of light rays reflected or refracted by a smooth curved surface
a caustic is a line or point where the light intensity approaches infinity when the wavelength tends to zero.
The Zeldovich Approximation (ZA) \citep{Zeldovich1970} in two-dimensional space is identical to refraction of 
parallel light rays by a  plate with thickness given by a two-dimensional  random smooth function. 
If a screen is placed behind the plate at a certain range of distances from the plate 
a bright pattern of caustics on the screen displays a two-dimensional web structure made of  'pancakes' and 'haloes'  
bounded by the caustics  \cite{Zeldovich1983}.  
 The intensity of light in caustic is high but of course finite because of the wave nature of light. 
 Similarly the density in caustics in a cold collision-less DM
is high but finite because the DM  is not a continuous medium and in addition due to a finite thermal velocity dispersion
\citep{Zeldovich_Shandarin:82}.    
Nevertheless the approximation of DM by a cold continuous medium is extremely  accurate and commonly used
in cosmology.

The multi-stream field is simply a count of the  streams with distinct velocities at every point in Eulerian space.
Generally the number of streams is an odd integer  except the caustic surfaces where it is an even integer.
The flip-flop field is the count of turns  inside out of each fluid element. It can be estimated  numerically either on particles
or tetrahedra of the tessellation of the three-dimensional phase space sheet that approximate the distribution of the cold
DM in six dimensional phase space \citep{Abel2012,Shandarin2012,Shandarin2014,Shandarin2016}. 

All three inherent  attributes of  the DM web — the multi-stream and flip-flop fields along with caustic surfaces — are  intimately 
related to each other  but they elucidate different aspects of the DM web 
 \citep{Ramachandra2015,Ramachandra17a,Ramachandra17b}.  
For instance the multi-stream field is naturally defined
in Eulerian space on arbitrary set of diagnostic points while  the flip-flop  field is  defined in Lagrangian space
 on particles or  tetrahedra. Of course
the flip-flop field can be  mapped to Eulerian space but this will result in significant change of its mathematical properties:
being a field in Lagrangian space it becomes  a multi-valued function in Eulerian space which is more difficult
to deal with. 
Caustics are the interfaces between regions with different values of flip-flops in Lagrangian space and between regions 
with different values of the number of streams in Eulerian space.

 In principle  caustics can be identified as small regions with extremely high DM densities but it would require
N-body simulations with unfeasibly high mass resolution in Eulerian space for realistic cosmological simulations.
However since caustics separate the Lagrangian neighbor elements with different number of flip flops
 it is also possible to use the common faces of two neighboring tetrahedra with  opposite volume
signs as an approximation to the elements of the caustic surfaces.

Topological analysis and classification of all  generic types of caustics originating in a potential mapping of
a collision-less media in two and three dimensions was provided by Arnold (1982). An example of such
a mapping is the ZA.
However the analysis was based on the so called normal forms which roughly speaking 
are the minimal polynomials used as generators of singularities. In this form the results can be used only 
as a solid guideline for the much more strenuous analysis of realistic DM flows in cosmological simulations.
The first  analysis of the geometry and topology of the caustic structures in the frame of the ZA with 
smooth random initial perturbations was done by \cite{Arnold1982}, however it was limited to two dimensions.

For the recent  scrutiny of  the subject see \cite{Hidding2014} and \cite{Feldbrugge2018}.
However \cite{Hidding2014} showed the 3D caustics only in  the ZA simulation
and \cite{Feldbrugge2018} showed the maps of three bicaustics (Arnold's terminology) in 3D N-body simulations
but not the caustics. In simple terms, a bicaustic is the trace of an instantaneous caustic moving with time. 
For example the  red surface in Figure 1 in  \citep{Feldbrugge2018}  referred to  in the caption as  A3 (Arnold's terminology) 
'caustic feature'  is actually the trace of A3 singularity.  A3 singularity is always a curve at any instant of time.

The presence of caustics has been implied in most if not all modern cosmological N-body simulations.
For example \cite{More2015}  defined a {\it splashback} radius as the distance from center of the halo 
to the outermost closed caustic as an alternative to virial radius for defining boundaries of DM haloes.
They argued that the splashback  radius is a more physical halo boundary choice than the 
one based on a density contrast $\Delta$ relative to a reference (mean or critical) density.
In practice they did not identify caustics at all.  Instead they searched for a minimum
of the logarithmic slope of the spherically averaged density profiles of the haloes. Spherical averaging of  the density profiles may
enhance the robustness of the results but it imposes an  assumption that haloes can be well described as spherical configurations. 
Direct identification of caustics may make this assumption unneeded.

The studies of the properties of cosmic velocity fields in a collision-less medium especially the multivalued character of 
the flows easily reveal the discontinuities at caustics, e.g.  \citep{Shandarin:11,Hahn-etal:15} and many others.

This paper focuses on explicit identification of caustic surfaces in N-body simulations using the Lagrangian submanifold.
We will discuss the major difficulties and suggest some approaches to their solution. 

In Section \ref{sec:ZA}, we briefly discuss Zel'dovich approximation in the context of caustic formation. Section \ref{sec:Method} describes the details of our algorithm using the Lagrangian tessellation scheme. In Section \ref{sec:CaustHigh}, analyses involving caustic surface are done in combination using length scales and flip-flop information, and their significance in delineating caustics at various levels of non-linearity. We also characterize caustic surfaces in Lagrangian and Eulerian spaces.

\section{Zel'dovich Approximation and singularities}  
\label{sec:ZA}
In this section we introduce the concept of singularities in a cold continuous and collision-less medium.
All three requirements are necessary for the formation of  singularities. Cold dark matter (CDM) is an almost perfect
example of such a medium. We begin with a brief illustration by describing the evolution of CDM density field
using the Zeldovich approximation (ZA).
The ZA is an elegant analytical approximation to describe the non-linear gravitational evolution of cold collision-less and continuous medium. Technically it is a first order Lagrangian perturbation theory. However Zel'dovich suggested to extrapolate it to the beginning of the non-perturbative nonlinear stage and predicted the formation of caustics which are the boundaries of the first very thin multistream regions dubbed by him as `pancakes'. The ZA describes a dynamical mapping from the initial Lagrangian coordinates $\mathbf{q}$ to Eulerian positions $\mathbf{x}(t)$ at time $t$. In comoving coordinates, $\mathbf{x} = \mathbf{r}/a(t)$ (where $a(t)$ is the scale factor assuming normalization $a(z=0)=1$ and $\mathbf{r}$ are the physical coordinates of particles at present), the ZA takes the form:

\begin{equation} \label{eq:ZA1}
 \mathbf{x}(\mathbf{q}, t ) = \mathbf{q} + D(t) s(\mathbf{q}),
\end{equation}
where $D(t)$ is the linear density growth factor, and  the initial perturbation field $\psi(\mathbf{q})$ determines the vector field $\mathbf{s(q)} = - \nabla_q \psi(\mathbf{q})$. The potential $\psi(\mathbf{q})$ is proportional to the gravitational potential at the linear stage.
Conservation of mass implies $\rho(\mathbf{x}, t) d\mathbf{x} = \rho_0 d\mathbf{q} $, so the density field at $t>0$ in terms of Lagrangian coordinates is given as 

\begin{equation} \label{eq:J}
 \rho(\mathbf{q}, t) = \rho_0 \left( J \left[ \frac{\partial\mathbf{x}}{\partial\mathbf{q}} \right]\right)^{-1},
\end{equation}
where the Jacobian $J \left[ \frac{\partial\mathbf{x}}{\partial\mathbf{q}} \right]$ is calculated by differentiation of  Equation \ref{eq:ZA1}. Moreover, diagonalization of the symmetric deformation tensor $d_{ij} = - \nabla_q \mathbf{s(q)} =  \partial^2 \psi(\mathbf{q})/ \partial q_i \partial q_j$ in terms of its eigenvalues $\lambda_1(\mathbf{q})$, $\lambda_2(\mathbf{q})$, $\lambda_3(\mathbf{q})$ particularizes the patterns of collapsing 
of the fluid elements. This reduces the equation describing the  mass density to a convenient form:

\begin{equation} \label{eq:density}
 \rho(\mathbf{q}, t) = \frac{\rho_0}{ \left[1 - D(t) \lambda_1(\mathbf{q}) \right]\left[1 - D(t) \lambda_2(\mathbf{q}) \right]\left[1 - D(t) \lambda_3(\mathbf{q}) \right] }.
\end{equation}


Since the deformation tensor $d_{ij}$ and its eigenvalues depend only on the initial  fields, the ordered eigenvalues defined in Lagrangian space $\lambda_1(\mathbf{q}) > \lambda_2(\mathbf{q}) > \lambda_3(\mathbf{q})$ determine collapse condition for fluid elements in Eulerian space. In the context of this paper, formation of caustics is of much interest: with increasing $D(t)$, the mass density of cold continuous 
fluid can rise until reaching singularity at $D(t) = 1/\lambda_1(\mathbf{q})$.  

Locally in Lagrangian space, the condition $\lambda_1(\mathbf{q})  = 1/D(t)$ firstly  takes place
at a maximum of $\lambda_1(\mathbf{q}) = max = 1/D(t_{0}) $  where $t_{0}$ denotes the time of the `pancake's birth'.
After that at $t > t_0$ they are the level surfaces where $\lambda_1(\mathbf{q})$ 
takes a given constant value $\lambda_1(\mathbf{q}) =1/D(t) < 1/D(t_0)$.  
At small $\delta t = t - t_0$ the level  surfaces are closed and convex in Lagrangian space.
In Eulerian space at time $t$ they form the surfaces where  density becomes formally infinite 
if the continuous  medium approximation  is used - therefore the term a caustic. 
Equation \ref{eq:density} shows that the maximum number of flip-flops experienced by a fluid element can be either three, two, one,  or zero 
if the eigen values of the element satisfy one of four conditions $\lambda_3 >0$, or  $\lambda_2>0$, or $\lambda_1 >0$, or  $\lambda_1 <0$ 
respectively.

Mapping of the interior of  such a surface into Eulerian space  by equation (\ref{eq:ZA1})  turns it inside out 
which is possible in the case of a collision-less medium.
The first collapse compresses such a region into  a very thin layer -- Zeldovich's pancake
which consists of three  overlapping streams moving with  distinct velocities.
The first eigen value is greater than the boundary value $\lambda_1(\mathbf{q}) >1/D(t)$ in the interior
region only and therefore has negative density according  to the formal application of equation \ref{eq:density}. 
This happens because the mapping from Lagrangian to Eulerian space results in a fold in phase space 
($\mathbf{x}(\mathbf{v})$) as well as in  $\mathbf{x}(\mathbf{q})$ space.
Thus the first pancakes are always the regions of three-stream flows  and bounded 
by a caustic surface.  The caustic surface separates the region with negative density from the
surrounding field of positive density in Lagrangian space while in Eulerian space it separates
the three-stream flow pancake from the  single-stream field surroundings. It is not surprising that 
the number of streams is  equal to  two on the caustic surface.


At a randomly chosen point  $\mathbf{q}$ the three eigen values are always  distinct from each other
if the perturbation $\psi(\mathbf{q})$  is a generic field.
Three fields  associated with three eigen values
are non-Gussian even if the generating potential  $\psi(\mathbf{q})$ is a Gaussian field. 
In the case of the Gaussian potential the joint PDF of three eigen values can be found in analytic form 
(see e.g. \cite{Doroshkevich1973} and \cite{Lee1998}). It contains a factor
 $(\lambda_1 - \lambda_2) (\lambda_1 - \lambda_3) (\lambda_2 - \lambda_3)$ which explicitly
  shows that the chance of finding a point with  two equal eigen values has a zero probability.

However it is worth stressing that  the points with two equal eigen values  exist but the  points with all
three equal eigen values do not. The points of equality of only two eigen values ($\lambda_1 = \lambda_2$ or 
$\lambda_2 = \lambda_3$) make up lines that are sets of measure zero in three-dimensional space 
thus the zero probability of finding them in a random search. The lines occur because the above equations
actually  hide two equations each. At the point of equality of two eigen values (say $\lambda_1 = \lambda_2$) 
two corresponding eigen vectors are degenerate in a plane that is orthogonal to the third eigen vector.  
This means that the corresponding minor in the deformation tensor is diagonal in any coordinates  in this plane,
which can be satisfied only if $d_{11} = d_{22}$ and $d_{12}=d_{21} = 0$.

A claim  $\lambda_1(\mathbf{q}) = \lambda_2(\mathbf{q}) = \lambda_3(\mathbf{q})$)  actually requires to satisfy 
five conditions simultaneously  and therefore 
the overdetermined system for three unknown coordinates $\mathbf{q}$  has no solutions. 
If such a point $\mathbf{q}(\lambda_1=\lambda_2=\lambda_3)$ existed than in arbitrary Cartesian system  six
components of a symmetric deformation tensor $d_{ik}= \partial^2{\psi} / \partial{q_i}\partial{q_k}$ 
at $\mathbf{q}(\lambda_1=\lambda_2=\lambda_3)$  must satisfy five equations: 
  $d_{11}=d_{22}=d_{33}$ and $d_{12}=d_{13}=d_{23}=0$ which is impossible in generic fields.

\begin{figure}  
\centering\includegraphics[width=9cm]{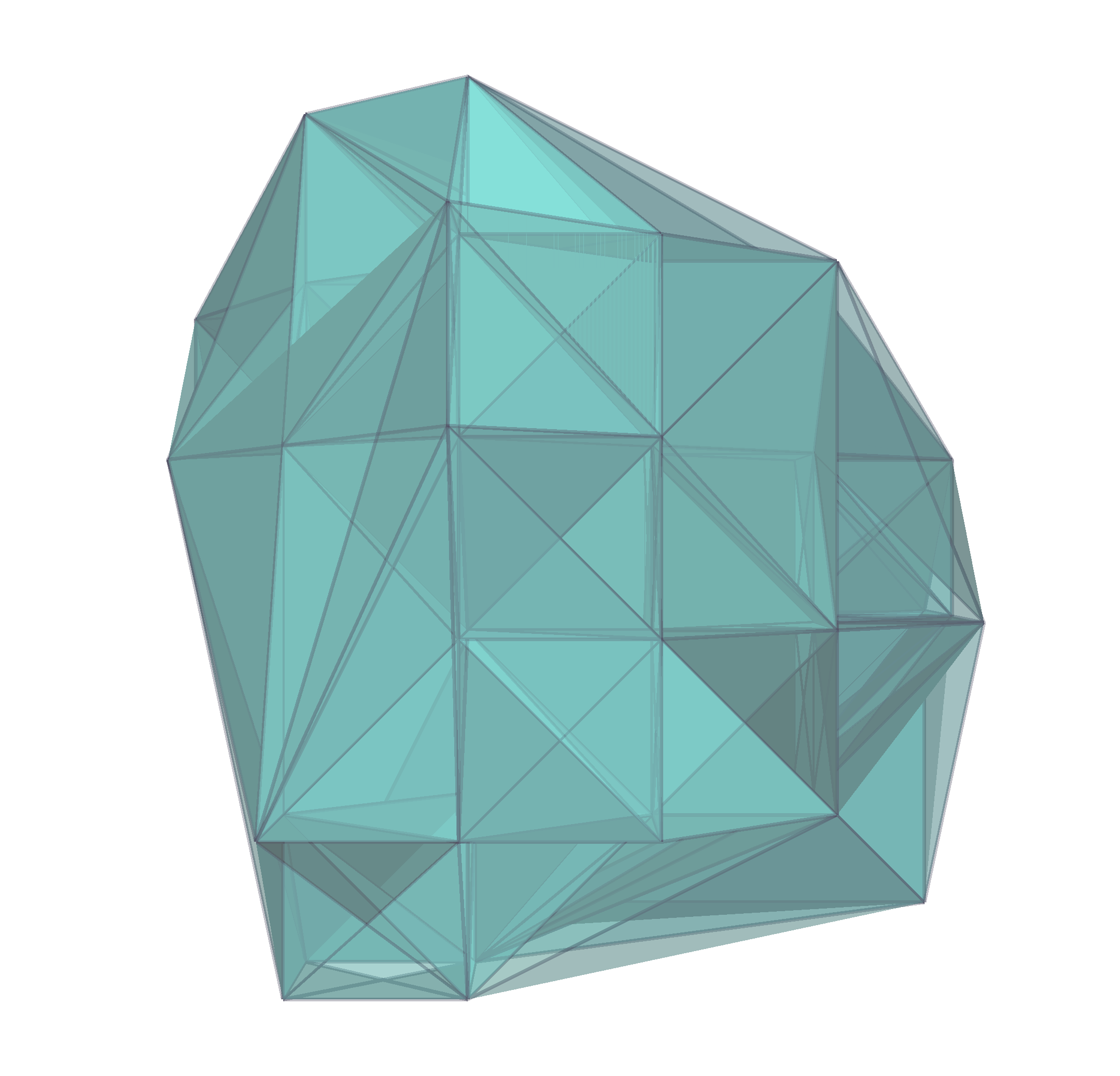}
\centering\includegraphics[width=5cm]{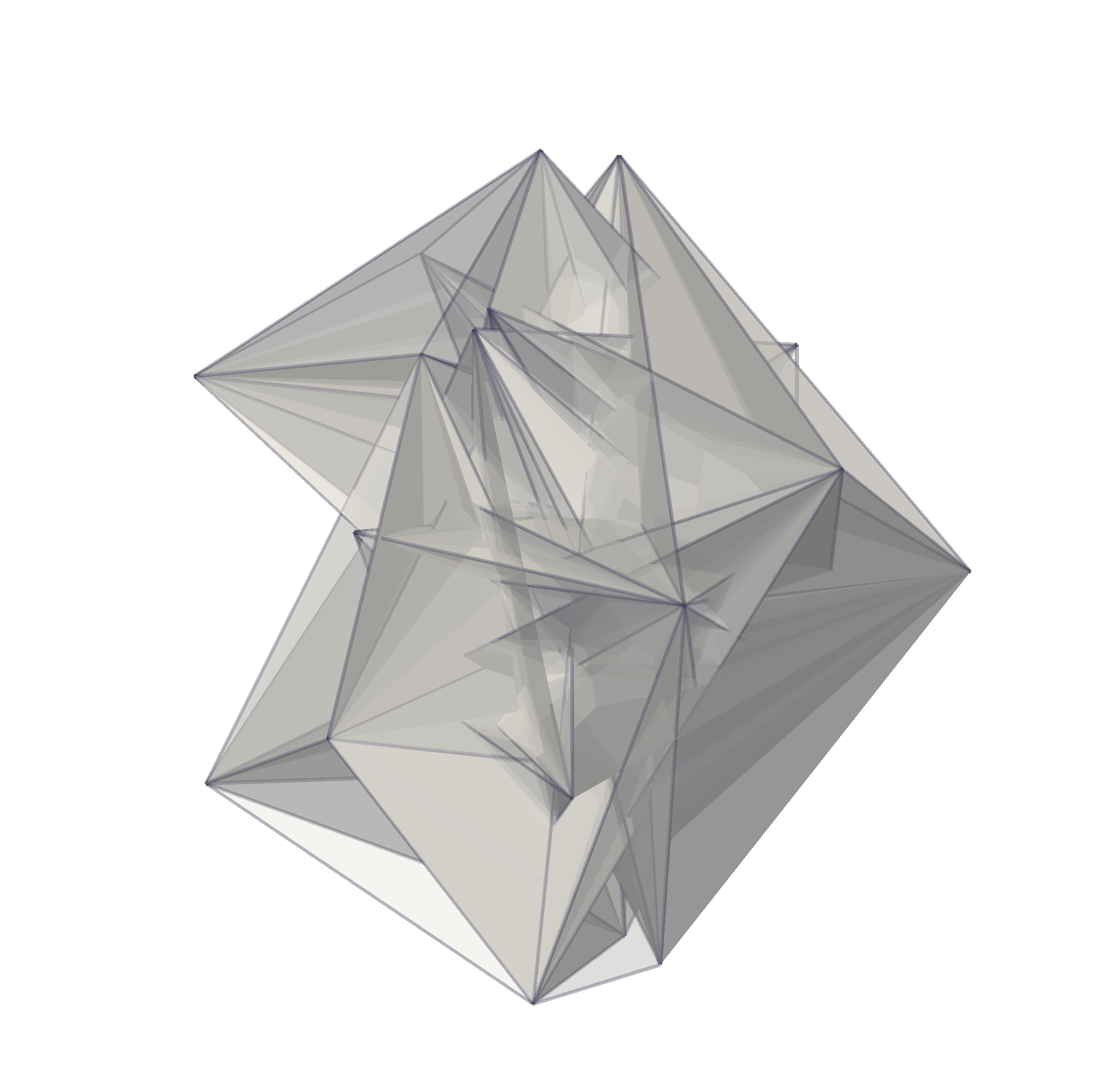}
\caption{ An  illustration of  the idea of Lagrangian tessellations. Left: The distribution of Dark matter particles in Lagrangian space is on the regular grid. The tetrahedra surfaces in this case are mostly regular (except at the edges). The bottom panel shows the Eulerian positions of the same particles at $z=0$. The particles clearly have undergone multiple flip-flops, and the intersections of Lagrangian tetrahedra signifies locations of caustic surfaces. Specific tessellation schemes can be utilized separately to identify these surfaces. }
\label{fig:Tess}
\end{figure}


For more detailed analysis of the geometry and topology of the caustic structures in  two-dimensional case we
refer to \cite{Arnold1982} and \cite{Hidding2014}.  
Unfortunately a detailed analytical characterization of 3-dimensional ZA with generic initial perturbations has not reach 
a similar level yet,  however an important step forward was made recently  by  \cite{Feldbrugge2018}.


The geometrical and topological complexities of the caustic surfaces  in three-dimensional configuration space 
 are due to the maze-like map of the three-dimensional hypersurface $\mathbf{x}(\mathbf{q})$ called the Lagrangian submanifold from six-dimensional $(\mathbf{q},\mathbf{x})$-space to three dimensional $\mathbf{x}$ space (See Figure \ref{fig:Tess}). The Lagrangian submanifold is a single valued, smooth and differentiable vector function 
 $\mathbf{x}=\mathbf{x}(\mathbf{q})$. However its inverse i.e. $\mathbf{q}=\mathbf{q}(\mathbf{x})$ is generally 
 multivalued vector function similar to $\mathbf{v}=\mathbf{v}(\mathbf{x})$.
 The projection onto 3-dimensional Eulerian space is entangled with creases, kinks and folds. Note that this  submanifold is very different from the phase space 
$(\mathbf{x},\mathbf{v})$, even though they are connected by a canonical transformation. Delineating the Lagrangian submanifold reveals several properties of the dark matter dynamics not inferred from position-space analyses. 
Two physically  related fields associated with tessellating the Lagrangian submanifold -- the multistream field $n_{str}(\mathbf{x})$ in Eulerian space \citep{Shandarin:11,Shandarin2012, Ramachandra2015,Ramachandra17a,Ramachandra17b} 
and the Flip-Flop field $n_{ff}(\mathbf{q})$ in Lagrangian space  
\citep{Shandarin2016} 
are instrumental for the analysis of the Lagrangian submanifold.

\section{Identification of caustics in numerical simulations} 
\label{sec:Method}
Finding caustics in numerical simulations by the ZA as well as by N-body  could be 
made by various methods.  One can do it by analyzing singularities of  the eigenvalue fields 
of the deformation tensor $\partial s_i /  \partial q_j$,
where $s_i(\mathbf{q},t) = x_i(\mathbf{q},t) - q_i$. In the case of ZA it is simply $D(t)s_i(\mathbf{q})$ 
in equation \ref{eq:ZA1}.
In the case  of N-body simulations  it requires numerical calculation of the positions of the particles at time $t$.
The passage of a particle through the singular stage can be  registered through the change of sign
of the Jacobian $J$ (equation \ref{eq:J}), see e.g.  \cite{Vogelsberger2011, Shandarin2016}.  
This method  allows to count the number of times a particle passed through the caustic state which is equivalent
to the number of  flip-flops computed for each particles.
 The probe of the geometry and topology of caustics
requires to analyze the spacial structure of the eigenvalue fields  \cite{Arnold1982,Hidding2014, Feldbrugge2018}, 
In this paper we will use a different method of finding caustic 
surfaces directly from the Lagrangian triangulation \citep{Abel2012,Shandarin2012}.
If two neighboring tetrahedra sharing a common face have different signs of volumes evaluated by the following equation 
then the common face is an element of a caustic or a cell of the triangulation of the caustic surface. 
The vertices of the triangles are obviously the points on the caustic. 
The caustic we identify by this method is A2 singularity  by Arnold's classification. All higher order singularities are
singular lines and point of the caustic surface A2, for instance cusps which are A3 singularities on curves.

The volume of a tetrahedron can be evaluated by computing the following determinant 
\begin{equation} 
\label{eq:det}
V = \frac{1}{3!}
\begin{vmatrix}
x_{1} & y_{1} & z_{1} & 1 \\ 
x_{2} & y_{2} & z_{2} & 1 \\ 
x_{3} & y_{3} & z_{3} & 1 \\ 
x_{4} & y_{4} & z_{4} & 1 
\end{vmatrix},
\end{equation}
where  $x_i, y_i, z_i$ are the coordinates of four vertices of a tetrahedron.
Of course the  vertices of all tetrahedra in the tessellation are assumed to be arranged in such an order 
that all volumes are initially positive. 
Like the Delaunay triangulation, the Lagrangian triangulation algorithms also consider particles as a tracers of collision-less DM fluid. 
However, as opposed to Delaunay triangulation scheme, this type of tessellation is carried out in Lagrangian space only once 
and remains intact during the entire evolution regardless of changes in the distances between the particles. 

The determinant of Equation  \ref{eq:det} is calculated for all tetrahedra at every output time but
not at every time step  of the simulation. Therefore the numbers of the tetrahedra  flip-flops are not computed.
However finding the triangle elements of caustics requires only the signs of the tetrahedra volumes
and therefore can be done without recording the history of flip-flopping.
We compute the flip-flops  on particles because it is easier than on tetrahedra and this is done at every time step of the simulation \citep{Shandarin2016}.

 We introduce two auxiliary characteristics of the cells/triangles:  the length of the longest side of the triangle $l_{max}$ and
the mean number  of flip-flops computed on the vertices of the caustic triangle $\overline{n_{ff}}(q_c)$ 
\begin{equation}
\label{eq:mean_ff}
l_{max} \equiv max(l_1,l_2,l_3)\,\,\,\,  {\rm and}\,\,\,\,   \overline{n_{ff}}(q_c) \equiv {1\over3}\sum_{i=1} ^3 n_{ff} (v_i),
\end{equation}
where $l_1,l_2,l_3)$ are the lengths of a caustic triangle and $n_{ff} (v_i)$ is the number of flip-flops on vertex  "i".


Our Lagrangian tessellation decomposes each elementary cube in five tetrahedra \citep{Shandarin2012}. 
In order to match the faces of  tetrahedra in two neighboring elementary cubes we use two types of the orientation of even and 
odd cubes. The tessellation involves two kinds of tetrahedra: the central one 
with volume $V_c=a^3/3$ and four corner tetrahedra with volumes $V=a^3/6$ where $a=L/N$ is the size of an elementary cube.
The central tetrahedron is regular with edges  $e_c = \sqrt{2} \,a$.  
Four equal corner tetrahedra have three edges of length $e_1=a$ and three  
of length $e_2=  \sqrt{2} \,a$. 

We  carried out  our simulations  with the standard $\Lambda$CDM cosmology, $\Omega_m=0.3,~\Omega_\Lambda=0.7,~ \Omega_b=0, ~\sigma_8=0.9,~ h=0.7$. We used a modified version (for details see \cite{Shandarin2016}) 
of a publicly available cosmological TreePM/SPH  code GADGET \citep{Springel:05}. 
In order to have sufficient particle sampling  we introduce an artificial  spherically symmetric sharp cutoff scale 
in the initial spectrum of perturbations which is a standard option in GADGET. 
Thus the simulations differed from each other by four  parameters: the size of the box $L$ in units of $ h^{-1}$ Mpc,
the box size $N_{box}$ (number of particles $N_p=N_{box}^3$), the force  softening scale $R_s$ in units of $ h^{-1}$ Mpc, 
and a cutoff scale  $n_{c}$ defined in GADGET by the equation   $k_{c} =  ( 2 \pi /L) (N_{box} / n_{c})$ meaning that 
the initial power spectrum $P(k) = 0$ for all $k >k_{c}$.

We tried several sets of the parameters of the simulations described above before we were able to reliably identify the caustics.
The simulation has the following parameters: $L = 100 h^{-1}$ Mpc, $N_p=256^3$, the force softening scale $R_s=0.8h^{-1}$ ~Mpc,
and the cutoff scale parameter $n_{c}=64$ which means that the initial spectrum covers a very small range from $k_{min}= ( 2 \pi /L)$ to 
$k_{max}= 4( 2 \pi /L)$. It means that the reliable identification of several internal caustics requires very large number of sampling particles.
Thus this simulation pretends only to describe the caustic structure  evolved  from a rather simple but still totally generic type of 
initial condition  in quite accurate gravitational field. In other words we do not infer other significant cosmological outcomes except
that the internal caustic structure cannot be reliably obtained from common N-body simulations.
Our choice of the cutoff scale roughly correspond to the scale of massive clusters of galaxies. 
Very roughly it might qualitatively illustrate the formation of clusters of galaxies in the HDM scenario
but definitely  not in CDM model.

\begin{figure} 	
\centering\includegraphics[width=9cm]{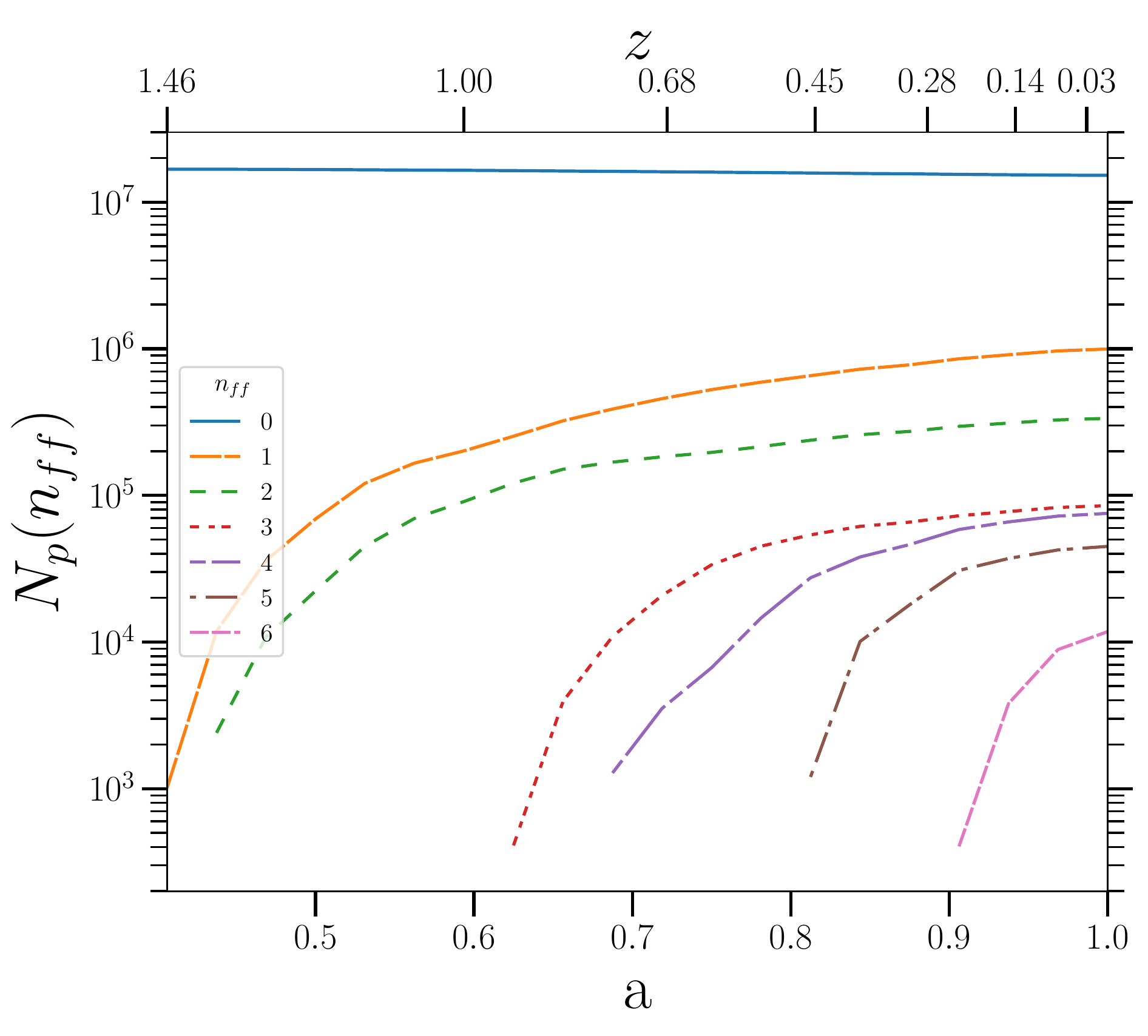}
\caption{Number of particle $N_p$ that experienced $n_{ff}$ flip-flops as a function of scale factor $a$. }
\label{fig:ff-evolution}
\end{figure}

\section{Caustics in a high sampling simulation} 
\label{sec:CaustHigh}
Figure \ref{fig:ff-evolution} provides a sense of the evolution of the structure in the simulation at the nonlinear stage.
 It shows the growth of the number of particles experienced flip-flops and the decrease of the number of particles
 with zero flip-flops.
We will discuss only the final stage of the simulation corresponding to $a=1$ and focus on the most evolved part of the box.

\begin{figure} 	
\centering\includegraphics[width=9cm]{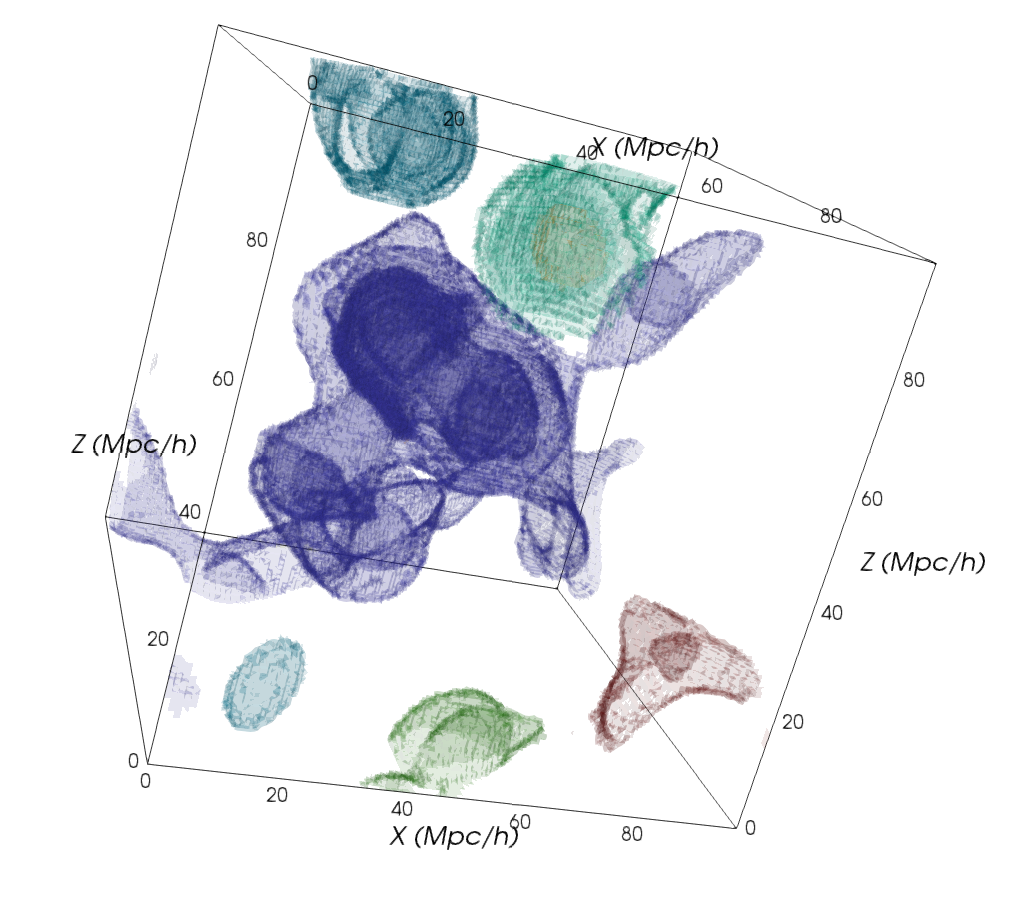}
\caption{Caustic surfaces in Lagrangian space. Darker regions are interior caustic surfaces.  
 The hues distinguish the isolated structures}
\label{fig:full-L}
\end{figure}

\begin{figure} 	
\centering\includegraphics[width=8cm]{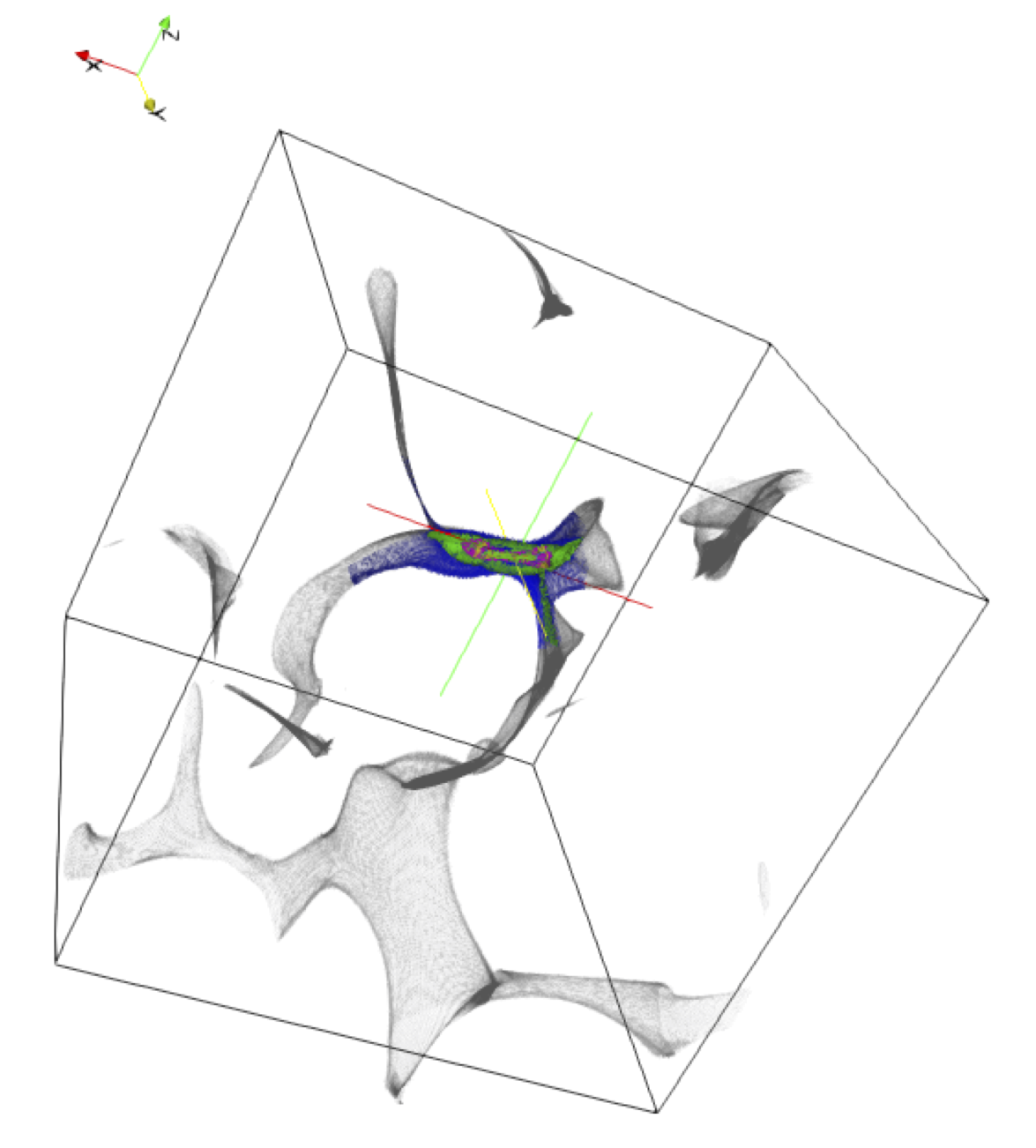}
\caption{Caustic surfaces  in Eulerian space. 
 The  colored area marks the region mostly evolved dynamically. It has been selected by a spherical clip with the radius of 19 Mpc/h }
\label{fig:full-E}
\end{figure}

The caustic surfaces in the  full simulation box are displayed  in Lagrangian space and Eulerian spaces in Figures \ref{fig:full-L}
and \ref{fig:full-E} respectively.
There are only a few largest caustic structures with more than a thousand of caustic elements i.e. caustic triangles/cells and 
caustic vertices/points.
Table 1 gives the numbers of triangular cells and  vertices in these structures. The geometry of the external caustic shells separating 
the multi-stream regions from the single-stream regions is typically simpler than that of the internal caustics in Eulerian space . 
The internal caustics are much more complex and their shapes practically has not been studied in cosmological N-body simulations. 
They are the major goal of our study.
The most dynamically advanced part of the caustic structure in this simulation is displayed in color  in  Figure  \ref{fig:full-E}.

\begin{table}{ Table 1. Number of caustic elements in  the structure shown in Figures \ref{fig:full-L} and \ref{fig:full-E}.}\\[1ex]
\begin{tabular} {l r r r r r r}
\hline
cells	&670310	&108955	&58988	&23905	&21138	&6886 \\
points &302999	& 51750	&28902	&11840	&10802	&3381 \\
\hline
\end{tabular}
\end{table}

\begin{figure} 	
\centering\includegraphics[width=9cm]{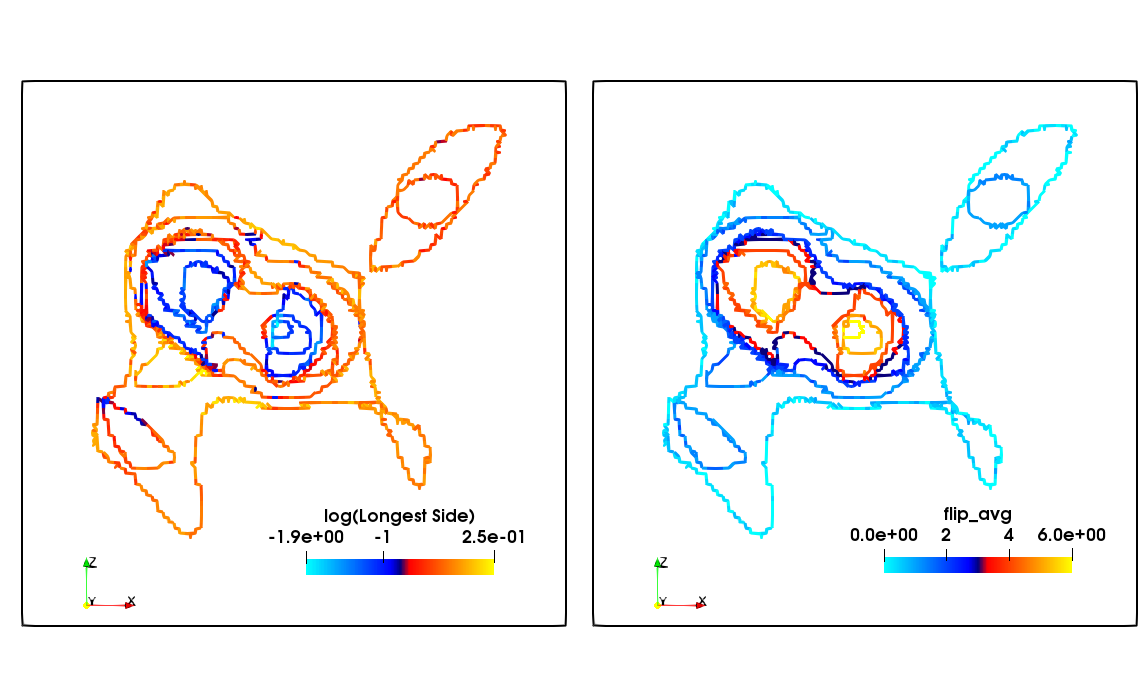}	
\centering\includegraphics[width=9cm]{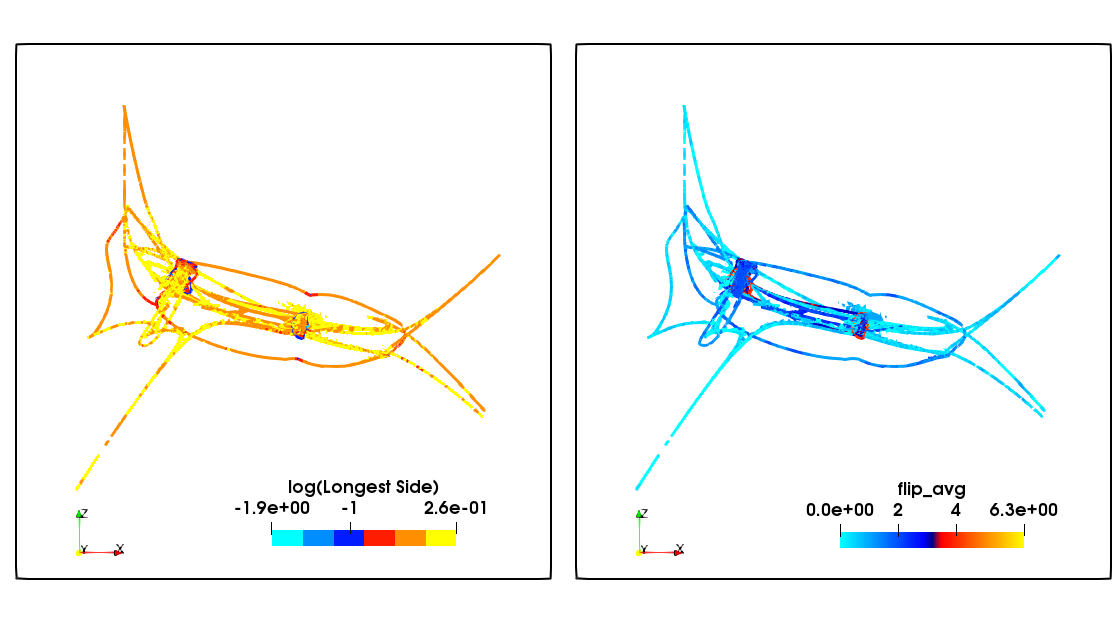}
\caption{The bottom panel: A razor thin slice through the colored region shown in Figure \ref{fig:full-E} in Eulerian  space.
The top panel: A razor thin slice through two haloes shown in Figure \ref{fig:center-L} in Lagrangian  space.}
\label{fig:center-EL}
\end{figure}

\subsection{Caustics in two-dimensional slices}          
First we demonstrate that the structures shown in Figure \ref{fig:full-E}  
are not the contours with arbitrary chosen levels as commonly used in cosmological N-body simulations but are the set 
of physical surfaces.
The  bottom panel of Figure \ref{fig:center-EL} displays the cross-section of the full set of the caustic triangles with a plane passing through 
two red bubbles clearly seen in Figure \ref{fig:full-E}.  Assuming that the set of independent triangles form  true surfaces one 
would conclude that the cross-section of a plane with this set of surfaces must be a set  of discrete  curves.  
Figure  \ref{fig:center-EL} unambiguously confirms this prediction. 
The top  panels  show the cross-section in Lagrangian space. 
It is worth stressing that the caustics in Lagrangian space provide much more clear idea of the structure
then in Eulerian space especially in dense crowded regions.

 In two panels on the left-hand side the colors of the curves code the logarithm of the longest side of the caustic triangles 
 crossed by the plane. And on the right-hand side panels colors show the mean number of flip-flops on the vertices of 
 the caustic triangles crossed by the plane. The color bars show the ranges of the caustic triangle sizes and the mean
 number of flip-flops. Both panels demonstrate a noticeable correlation of extended triangles with small counts of flip-flops.
 However the correlation of small size triangles with large counts of flip-flops is clearly demonstrated only
 in  Lagrangian space. One can see that the parts of the curves corresponding to small triangles  are 
 covered by the parts corresponding to extended triangles in both bottom panels by magnifying Figure  \ref{fig:center-EL}.

 The arrangement of the caustics in  Eulerian plane at the bottom of Figure  \ref{fig:center-EL} is substantially 
 more cumbersome than that in Lagrangian plane in the top panels. It is not surprising because the caustic surfaces can and do 
 have multiple intersections with themselves in Eulerian space  \citep{Arnold1982,Hidding2014}. The two-dimensional plots
 of caustic cross-sections in the top panels of Figure \ref{fig:center-EL} comprehensibly suggest that the caustic surfaces 
 are closed contours in Lagrangian space. 
 All parameters of the particles or caustic triangles including number of flip-flops and the maximal length of
 the caustic triangles are fields in Lagrangian space since they are  single-valued functions. Therefore understanding 
 the complexity of the caustic surfaces inside dense parts  of the cosmic web,  i.e. haloes, is considerably less demanding
in Lagrangian space  than that in Eulerian space.

 \begin{figure} 	
\centering\includegraphics[width=8cm]{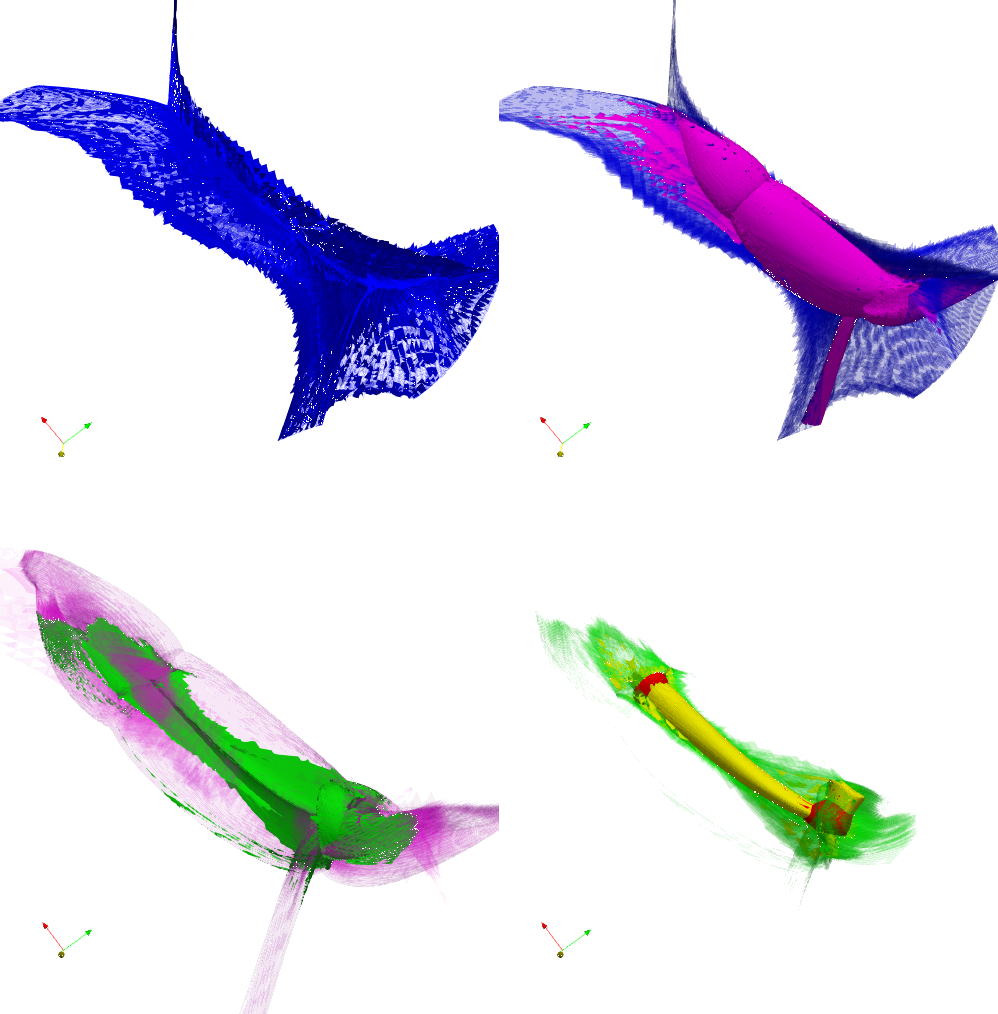}
\caption{The colored part of Figure \ref{fig:full-E} is magnified and shown in four panels. Five caustic shells can
be seen: blue, magenta, green, yellow and red. The radius of a spherical clip is around 19 Mpc/h.
However the bottom panels have been zoomed in and the distance between red bubbles is around 5 Mpc/h.}
\label{fig:center-E}
\end{figure}
\begin{figure} 	
\centering\includegraphics[width=7.cm]{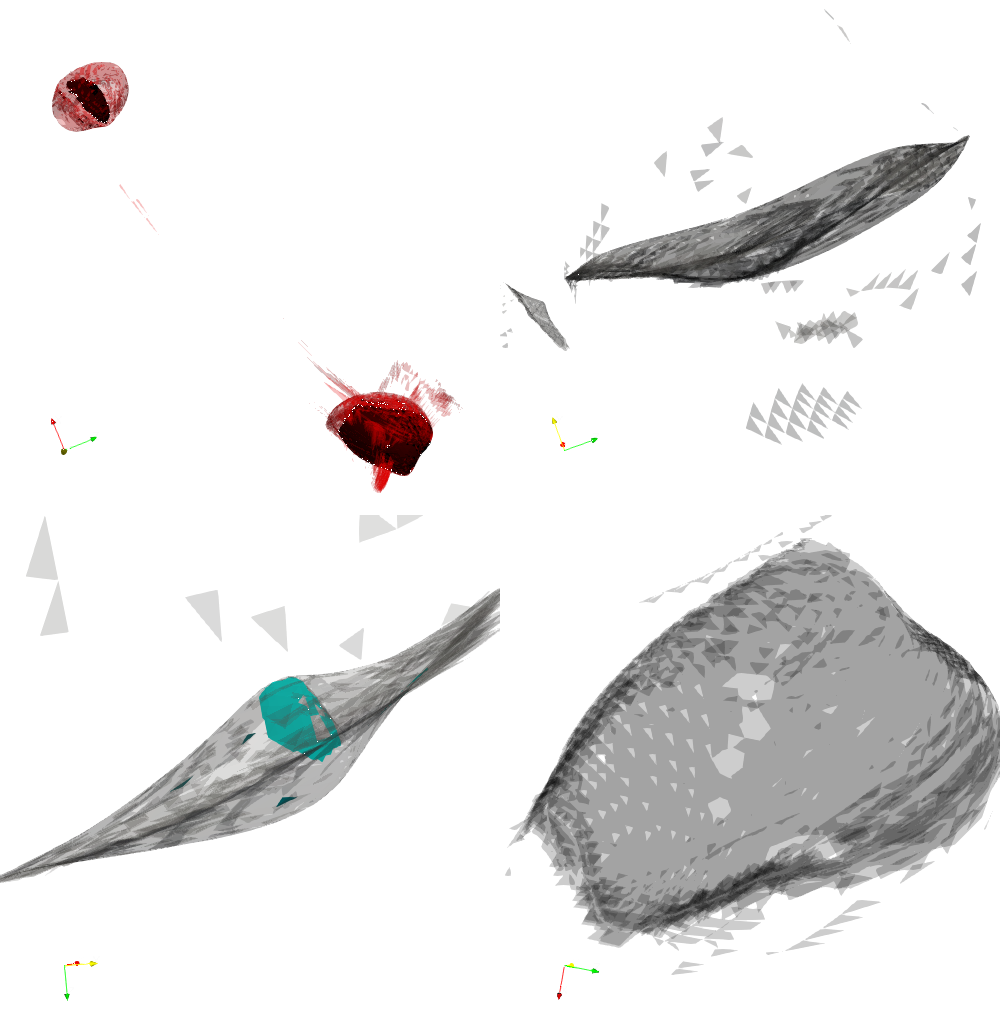}
\caption{The red bubble  like caustics are same as in Figure \ref{fig:center-E}.  
Black  caustics inside red bubbles are magnified and shown in gray.
Two  images on the right hand side show the same caustic from the lower red bubble in two approximately orthogonal  projections.
It can be  encircled by a sphere of radius 1 Mpc/h and its thickness is around  0.15 Mpc/h.
The gray caustic on the left hand side is from the top red bubble. 
 It can be  encircled by a sphere of radius 0.65 Mpc/h and its thickness is 
 around  0.15 Mpc/h.  It contains one more caustic (cyan) oriented in orthogonal direction to the  parent caustic. }
\label{fig:A6}
\end{figure}
\begin{figure} 	
\centering\includegraphics[width=9cm]{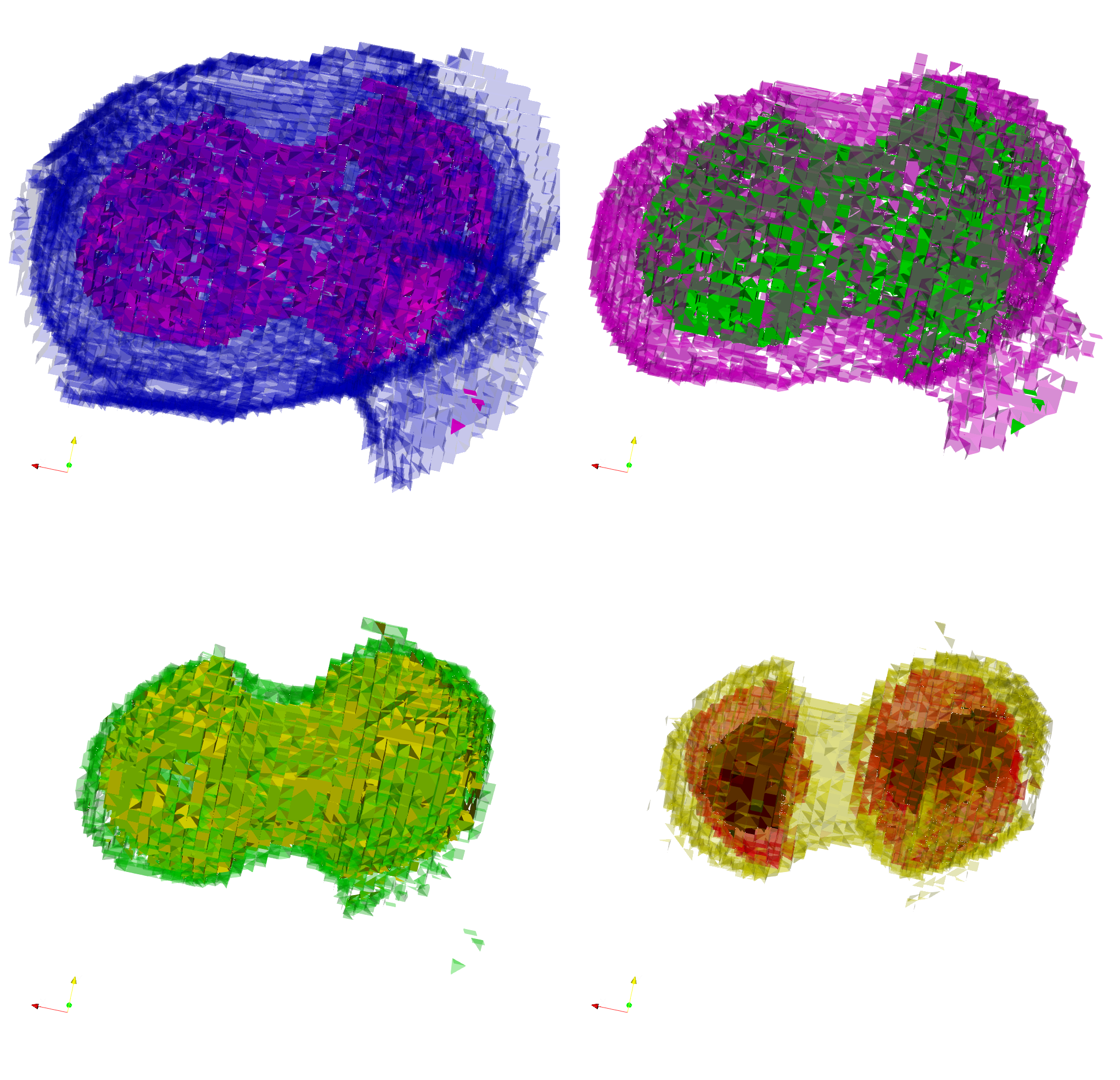}
\caption{The caustics in Lagrangian space corresponding to Figure \ref{fig:center-E}.
The radius of a spherical clip is around 26 Mpc/h.}
\label{fig:center-L}
\end{figure} 

\subsection{Shapes of caustics in three-dimensions}    
Now we will discuss in more detail the shapes of the internal caustics in three dimensions.
First of all it is worth stressing that the caustics are truly real physical objects although  up to accuracy of the physical model.
Therefore  caustics must be  unmistakably distinguished from contour plots of density or other fields.
A distribution of caustics   in space  represents a specific intermittent phenomenon in the sense that  it is a physical system
that has two states: one is the caustic surfaces and the other empty space. Alternatively it can be considered as a purely
geometric structure made up from two-dimensional surfaces. Both require very specific methods of analysis.
The caustics change their positions with time but physical velocities cannot be assigned to their triangular elements.
It is worth stressing that caustics move with phase velocities.
 We  provide visualization of several examples some of which display unexpectedly bizarre shapes.
 
 As we mentioned earlier the caustic surfaces consist of mutually independent triangles each of which was fully defined 
by two neighboring tetrahedra. It is totally independent  of other triangles in the caustic surface. 
Figure \ref{fig:center-EL} suggests that setting thresholds on 
$\overline{n_{ff}}(q_c)$ (equation \ref{eq:mean_ff}) may help to isolate particular parts of the total caustic surface.
As we demonstrate in Appendix A each pattern displayed in Figure \ref{fig:center-E}  and \ref{fig:A6} is  specified by  three or four 
components selected by a single  value of $\overline{n_{ff}}(q_c)$ (Table 2 ).
The figures in Appendix A1 and Figure  \ref{fig:A6} display all the components of  each  of seven caustic shells.
 
The colored patch  in  Figure \ref{fig:full-E}  is magnified and dissected into seven distinct patterns 
displayed  in Figures \ref{fig:center-E} and \ref{fig:A6}. 
Figure  \ref{fig:center-E} shows five caustic shells starting from two  external  shells in blue and magenta in two top  panels.
Two bottom panels display internal shells in green, yellow and  red.
 Figure \ref{fig:A6} shows the deepest caustics in gray  and cyan inside of two red bubbles displayed  in  Figures \ref{fig:center-E}.
 In both figures we try to show that green caustic is inside of the magenta shell and yellow and red one are inside of the green caustic.

 The top right panel in Figures \ref{fig:center-E} shows the blue and magenta   caustics crossing each other. 
The blue  caustic is painted with low opacity which allows to see more of the  caustic in magenta.
The middle part of the  caustic  in magenta approximately looks like a cylinder or tube. The  cross section plane passing through its axis
gives the impression of its three-dimensional contour (see the bottom of  Figure \ref{fig:center-EL}).
Then in the bottom left panel the caustic in magenta is plotted with low opacity which allows to see the internal caustic in green. 

The next tube - like caustic in yellow is shown in the bottom right panel. It is inside of the green caustic. 
In the cross section perpendicular to its axis the yellow caustic looks like almost perfect circle. 
Two red caustic are compact closed surfaces although not perfectly spherical but there is no considerable difference of 
their orthogonal diameters.  Their internal structures of red caustics are dissected in  Figure \ref{fig:A6}.

The top red bubble can be  encircled by a sphere of radius 0.9 Mpc/h
while the bottom red bubble is a little greater and can be placed inside a sphere of radius 1.1 Mpc/h. 
They are separated by the distance  around 5 Mpc/h. 
Two  gray images on the right hand side show the same caustic from the lower red bubble in two approximately orthogonal  projections.
It can be  encircled by a sphere of radius 1 Mpc/h and its thickness is around  0.15 Mpc/h.
We are suggesting that the red caustics would be good candidates for the outermost convex caustic of the haloes
that could explain a phenomenon analogous to 'splashback radius' \citep{More2015, Chang2018} on cluster scales.

Figure \ref{fig:center-L} shows the same six caustic shells in Lagrangian space. The figure clearly 
demonstrate the nesting structure from black, red yellow in the bottom right panel to green, magenta and 
blue in the top left panel. It allows one to estimate the  masses within the caustic surface in Lagrangian space.
The radii of the smaller and larger gray caustic shells  in Lagrangian space are roughly 4.5  and 6 Mpc/h containing  $1.3 \times 10^{14}M_{\bigodot}$
and $3 \times 10^{14}M_{\bigodot}$ respectively. 
While the radii of the smaller and larger red caustic shells  in Lagrangian space are roughly 6.5  and 9 Mpc/h containing  $4 \times 10^{14}M_{\bigodot}$
and $ 10^{15}M_{\bigodot}$ respectively. The estimates are quite crude because the shapes of the caustic surfaces in Lagrangian space are are not exactly spherical.

\section{Summary and Conclusion}  
\label{sec:summary}
This is the first successful attempt to directly build the caustic surfaces in three-dimensional N-body simulation. This required  a rarely used large force softening scale which was about twice of the mean particle
separation. Even more unusual was a large cutoff scale in the initial density perturbation power spectrum.
The wavelength of the cutoff was sixty four times of the mean particle separation scale or thirty two
Nyquist wavelength.  With such a small range of the remaining part of the power spectrum  the results
have little direct cosmological significance. However the results demonstrate that caustic structure 
formed in self gravitating collision-less medium from generic initial perturbations is very different  
from that in the case of the Zeldovich approximation. On the other hand the structure described  
in this paper is remarkably similar to Figure 4 in \citep{Melott1989}
obtained in the high resolution two-dimensional N-body simulation.

While the splashback radius \citep{Diemer2014, More2015} definition of the halo boundary detects a sharp change in the spherical density profile, the inner caustics are completely disappeared in their analysis. The attempts to rectify for spherical approximation \citep{Mansfield2017} to find aspherical shells also rely on density fields which are noisy in the inner regions of the haloes. With a Lagrangian sub-manifold triangulation technique, here we have successfully identified the physical caustic surfaces without any halo-boundary definitions, density computation schemes or empirical thresholds. Resulting caustics display a rich variety of caustic surfaces that can be delineated layer-by-layer either based on gravitational collapse history of the dark matter particles via the flip-flop field or the length scales of the caustic triangles. The splashback surfaces may roughly correspond to one of the several convex caustic surfaces surrounding a halo, but they neither represent a physical boundary of a halo that is independent of empirical modeling nor sufficiently identify the internal caustic structures.

The caustics must be described from simple contours plots because they are physical phenomena. In contrast with contour plots, the positions of the caustics are strictly determined by the laws of dynamics. However caustics are not physical objects like particles. For instance they change the positions but it is due to phase velocity rather than physical velocity. Thus caustics are discrete structures eventuated in flows of continuous collision-less media. 

Mapping of caustics to Lagrangian space reveals that the caustic surfaces are better resolved than in the Eulerian space. This is not too surprising, given the particles in Lagrangian space are on the grid. However, it is noteworthy that the hierarchal arrangement of the caustic surfaces remains intact. That is, the inner caustic in the Lagrangian space corresponds to an inner caustic in Eulerian space, although the surface geometries tend vary considerably.

There are two conspicuous examples of caustics revealed by this simulation. Neither compact approximately round shells  nor  tube-like caustic connecting red bubbles  (Figure \ref{fig:center-E}) exist in the Zeldovich approximation. Since these types of structures have occurred in a statistically small
simulation then they must be quite common.
Each compact round red shell is likely an indication of the beginning
of a halo formation.  If it is confirmed by further larger simulations then it may become a method
of identifying haloes and substructures in cosmological dark matter simulations based  on physics.

The next unexpected event is the formation of a pancake - like structures (Figure \ref{fig:A6}) within
quasi spherical caustic shell. One of such structures even encompasses another also pancake-like
structure. It's orthogonal orientation with respect to the parent pancake would be typical for the secondary 
pancake formed inside  the primary
pancake since the principal axes of the deformation tensor (equation  \ref{eq:J}) are mutually orthogonal.

The reliable identification of simplest pancake-like requires at least a thousand caustic elements i.e.  vertices and triangles
(see Table 2 in Appendix A).
The smallest caustic shell built by  373 triangles and 387 vertices is barely seen in Figure \ref{fig:A6}.
The issue of insufficient sampling is not uniquely  associated with caustic identification. 
 Recently \cite{vandenBosch-etal18} showed that it is impossible to reliably trace the evolution of subhaloes 
 unless the subhalo has at least 1000 particles.  \cite{2018MNRAS.477.3230S} faced it in an attempt to estimate
 the median density in the Universe. By using face-space interpolation techniques
\citep{Abel2012,Shandarin2012,Hahn2016a,2016JCoPh.321..644S} they created $32^3$  times more re-sampled
particles to infer high-quality density field. We are expecting that a similar approach will also work for 
identification of caustics surfaces in dark matter N-body simulations.

\section*{Acknowledgements}
Sergei Shandarin acknowledges the support from DOE BES Award DE-SC0019474. Nesar Ramachandra's work at Argonne National Laboratory was partially supported under the U.S. Department of Energy contract DE-AC02-06CH11357.

\bibliographystyle{mnras}
\bibliography{library}
\begin{verbatim}


\end{verbatim}

\appendix
\section{Identifying distinct caustic structures}
A caustic element -- a caustic triangle and its vertices -- is defined as the shared face of a pair of neighboring tetrahedra 
having opposite signs of their volumes evaluated by equation \ref{eq:det}. Each caustic triangle is treated as independent element. 
Therefore the caustic surfaces can be completely determined by a local condition. The mean value of the flip-flop
counts on three vertices of a caustic triangle (equation \ref{eq:mean_ff}) is discrete:  $\overline{n_{ff}}(q_c)=n\times (1/3)$ because
the  number of counts of flip flops on vertices is integer.
The number of the caustic elements for each value of $\overline{n_{ff}}(q_c)$ 
is given in Table 2. A set of four figures is provided as an illustration of a distinct pattern of a caustic structure. 
We identified totally seven distinct pattern shown in six  Figures from A1 through A5 and in Figure \ref{fig:A6}. 
Figure A1 consists of four components: all triangles with  $\overline{n_{ff}}(q_c)$=0, 1/3 and  2/3 and all but the largest
connected set  with $\overline{n_{ff}}(q_c)$=1.  
Figure A2 also consists of four components: the largest connected
set of triangles with $\overline{n_{ff}}(q_c)$=1, all triangles with $\overline{n_{ff}}(q_c)$=4/3 and 5/3 and all but the largest
connected set  with $\overline{n_{ff}}(q_c)$=2.  
Figure A3 consists of three components: the largest connected set of triangles with $\overline{n_{ff}}(q_c)$=2
and all triangles with  $\overline{n_{ff}}(q_c)$=7/3 and 8/3.
Figures A4 and A5 consist of all triangles with $\overline{n_{ff}}(q_c)$  indicated in Table 2.
 Figure 7 is of a  different design: the top left panel in red is the same as the bottom right panel in magenta in Figure A5  except the black
 caustics corresponding to $\overline{n_{ff}}(q_c)$ = 5, 16/3 and 17/3 were added. The gray caustics are magnified versions of
 black caustics seen inside red caustic shells. Two images of the caustic from the lower red bubble are two orthogonal projections.
The gray caustic within the bottom left panel is a magnified version of the caustic in the left top red bubble. It contains
one more caustic in cyan.

It is easy  to visually select the components practically for every pattern.  We label six pattern by the tag of the
corresponding figure: A1, A2 etc. The smallest caustic shown in cyan in Figure \ref{fig:A6} is referred to as A7. 
Visually the patterns of the components match reasonably well but not perfect.  For instance A2 in blue contains some cells that definitely
belongs to A1, A3 blue contains a little part of A2. We leave the solving of this problem for the further work.

Figure 7 deserves a special comment. Two gray caustics  looks like  Zeldovich's pancakes placed inside of  two compact 
(quasi spherical) bubbles A5. This is quite unexpected in the frame of widely accepted  paradigm restricting the formation
of pancakes only to two circumstances. One of which is the very beginning of the nonlinear stage and the other when
larger scales reach the nonlinear stages. Here we observe the formation of a pancake inside of a roughly spherical caustic.
Moreover inside pancake A6 there is another much smaller pancake A7 (made only of 373 triangles and 387 vertices) oriented 
perpendicular to the parent pancake. This is known to happen in the Zeldovich approximation when the pancake corresponding
to the second highest eigenvalue forms inside the primary pancake related to the highest eigenvalue.
 

\begin{table}{ Table 2. The number of caustic elements for each value of $\overline{n_{ff}}(q_c)$ }
\begin{tabular} {l l r r r r r}
\hline
          &$\overline{n_{ff}}$ &0		&1/3		&2/3	   	&1	& total \\
A1      &N($\Delta)$	    &28439  	&41737		&27141	  	&5195	&78112\\
           &N(V)	                 &26306  	&40143		&32619	   	&8586	&107654\\
           &color	                 & blue 	&green		&red	  	&yellow	&\\
\hline
   &$\overline{n_{ff}}$	    &1		&4/3		&5/3		&2	& total	 \\
A2            &N($\Delta)$	    &51518     	&33484  	&21954		&3057	&110013\\
           		&N(V)	    &33360   	&31650  	&25916		&5317	&96243\\
            &color	                 & blue 	&green		&red	  	&yellow	&\\
\hline
&$\overline{n_{ff}}$	    &2		&7/3		&8/3 		&	&total	 \\
A3   &N($\Delta)$	    &49245 	&23261 	&14600		&	&87106\\
           &N(V)	                &29860   	&21813 	&16678		& 	&68351\\
             &color	                 & blue 	&green		&red	  	&	&\\
\hline
&$\overline{n_{ff}}$	    &3		&10/3		&11/3		&	& total	 \\
A4 &N($\Delta)$	    & 31365 	&14662		&5873 		&	&51900 \\
           &N(V)	                & 21638   	&14274		&7575		&	&43487\\
              &color	                 &blue  	&green		&red	  	&	&\\
\hline
&$\overline{n_{ff}}$	    &4		&13/3		&14/3		&	& total	 \\
A5   &N($\Delta)$	    &20270 	&7373		&2135		&	&29778 \\
           &N(V)	                &13340 	&7510		&3014		&	 &23864 \\
              &color	                 & blue 	&green		&red	  	&	&\\
\hline
&$\overline{n_{ff}}$	    &5		&16/3		&17/3		&6-7	& total	 \\
Fig. 7   &N($\Delta)$	    & 6411 	&3442		&859		&373	& 11085\\
           &N(V)	                 & 4753 	 &3588		&1368		&387	&10096 \\
            &color                 & gray 	&gray		&gray	  	&cyan	&\\
\hline

\end{tabular}
\end{table}

\begin{figure} 	
\centering\includegraphics[width=7.cm]{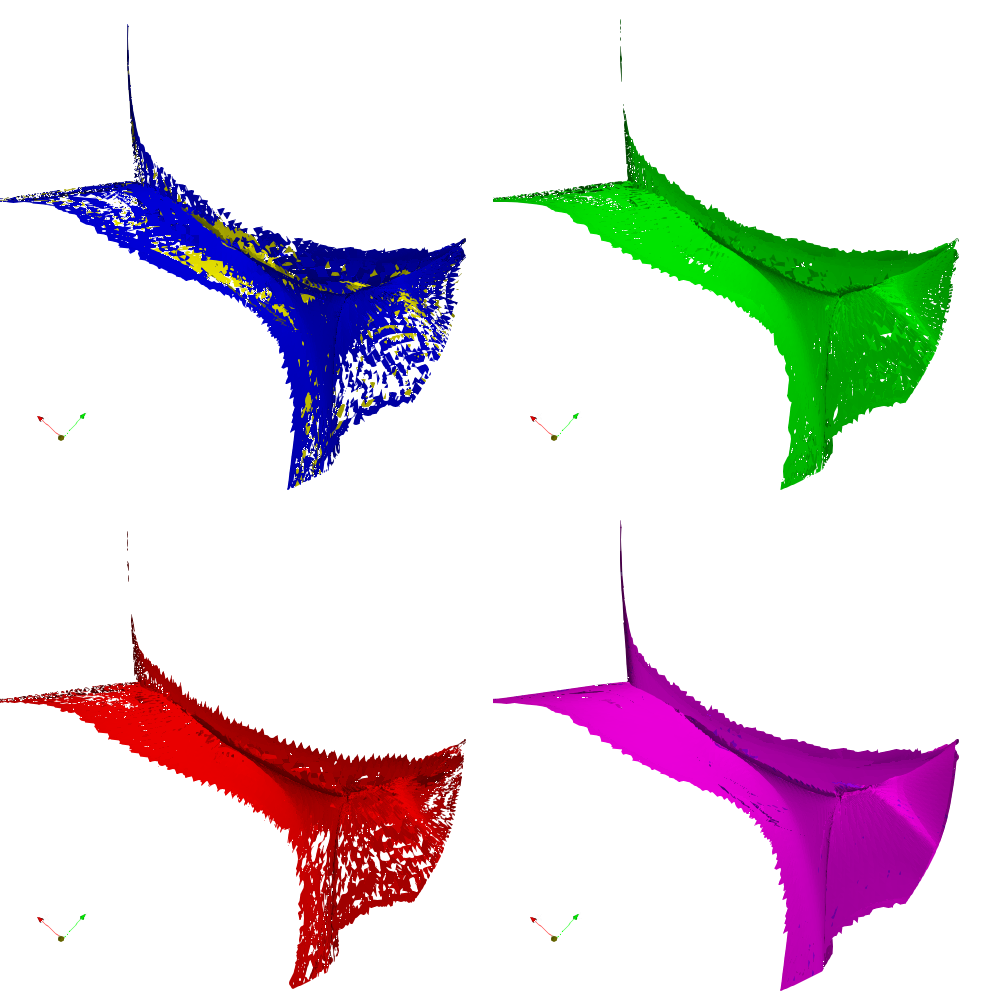}
\caption{ The caustic pattern (bottom right)  and its components (A1 in Table 2)}
\label{fig:A1}
\end{figure}

\begin{figure} 	

\centering\includegraphics[width=7.cm]{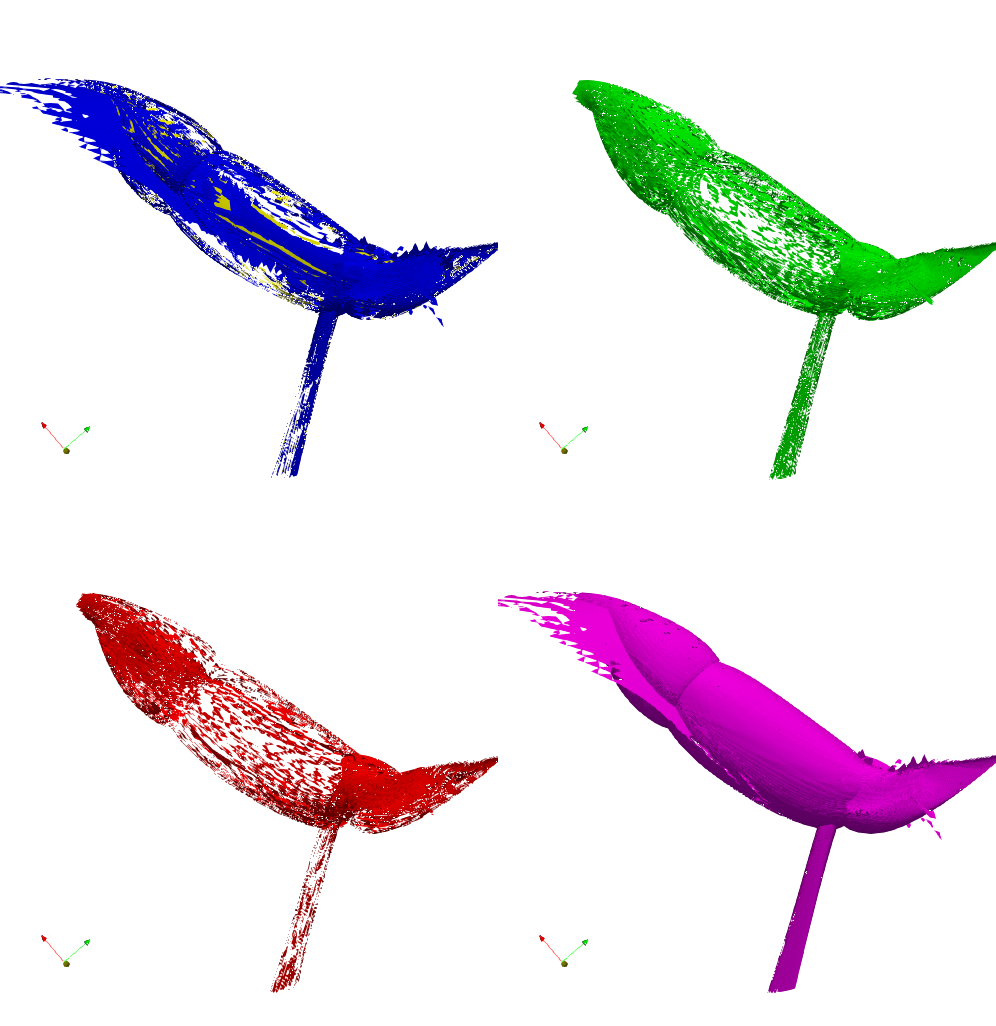}
\caption{ The caustic pattern (bottom right)  and its components (A2 in Table 2)}
\label{fig:A2}
\end{figure}

\begin{figure} 	
\centering\includegraphics[width=7.cm]{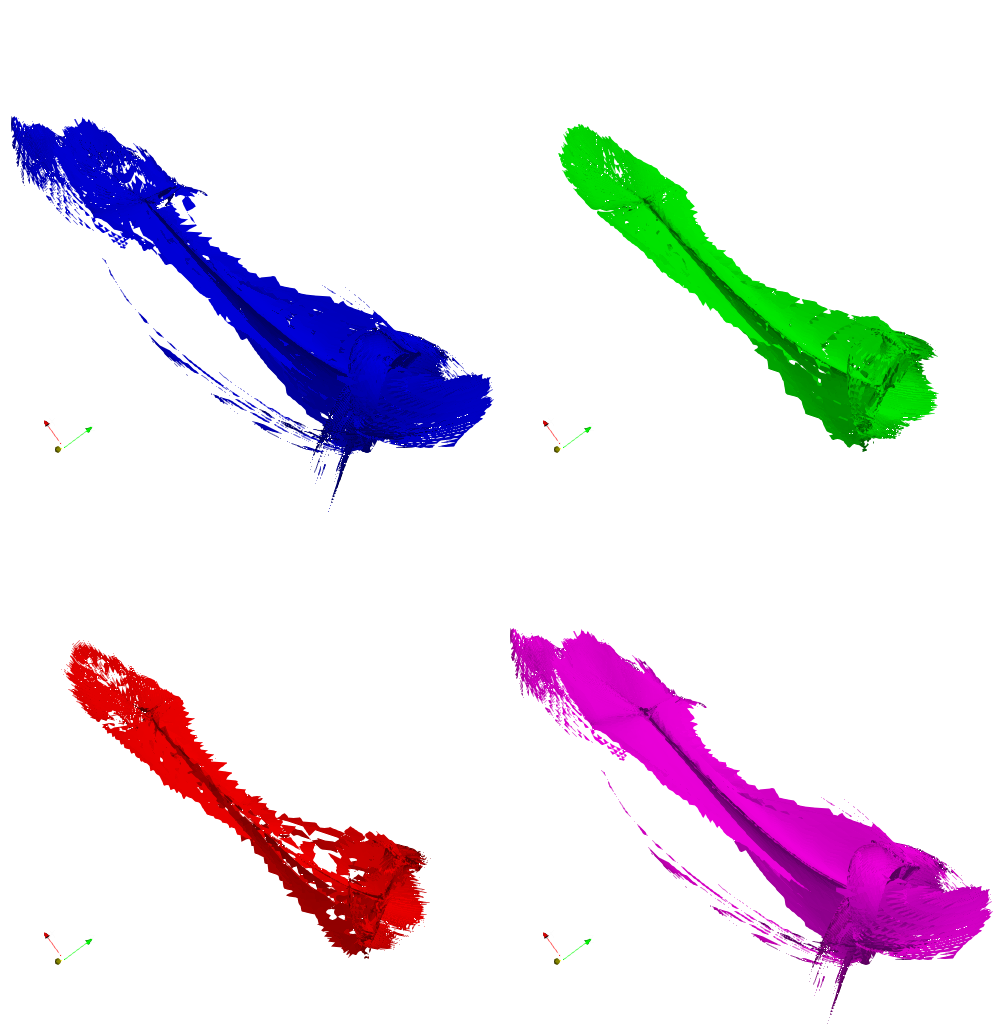}
\caption{ The caustic pattern (bottom right)  and its components (A3 in Table 2)}
\label{fig:A3}
\end{figure}

\begin{figure} 	
\centering\includegraphics[width=7.cm]{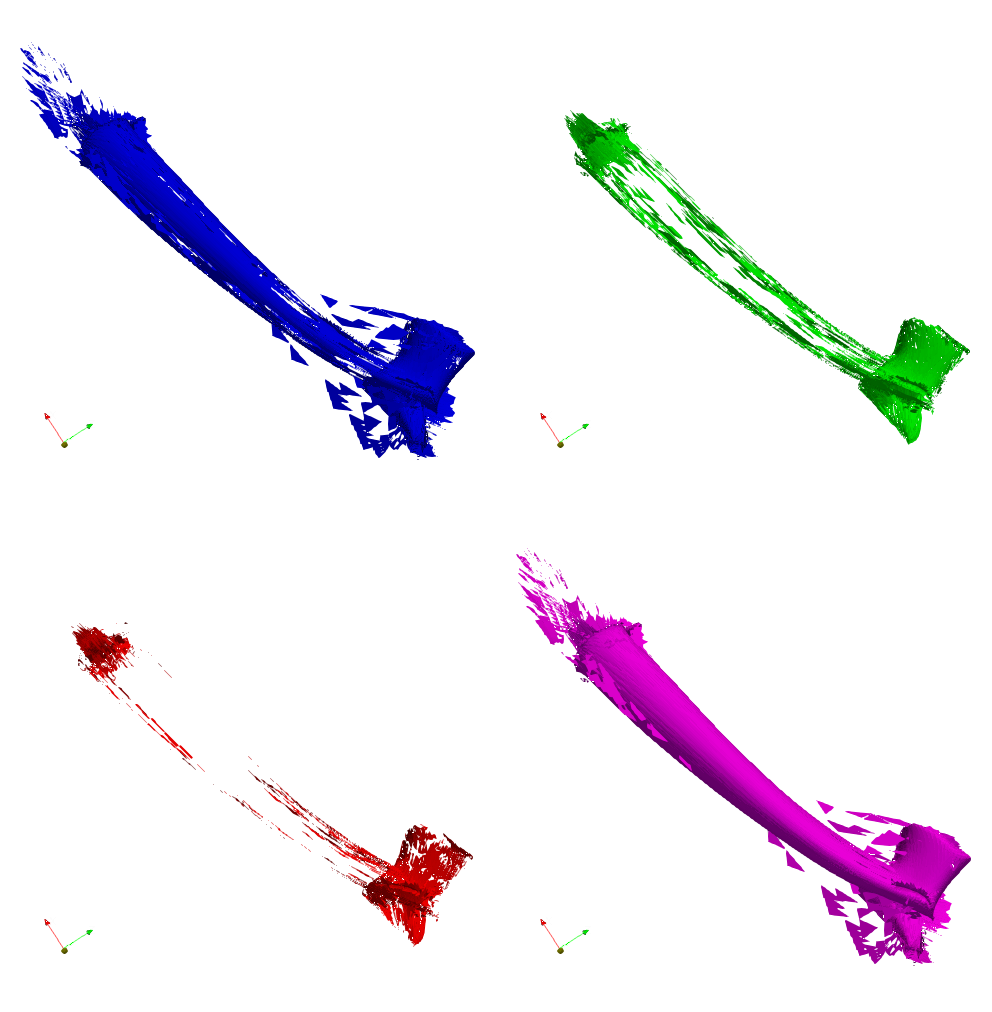}
\caption{ The caustic pattern (bottom right)  and its components (A4 in Table 2)}
\label{fig:A4}
\end{figure}

\begin{figure} 	
\centering\includegraphics[width=7.cm]{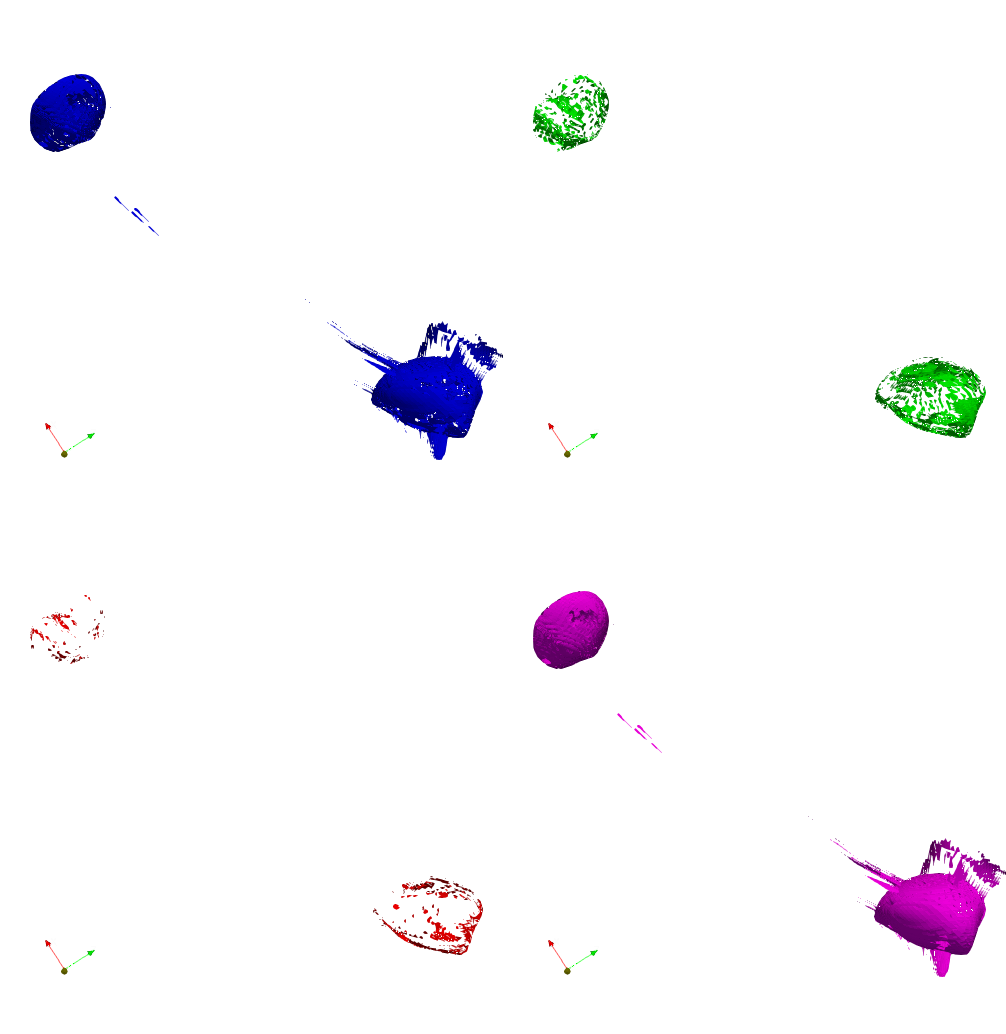}
\caption{ The caustic pattern (bottom right)  and its components (A5 in Table 2)}
\label{fig:A5}
\end{figure}

\label{lastpage}
\end{document}